\documentclass[a4paper]{jpconf}
\usepackage{graphicx}

\def\cpd{\rm{kg^{-1}keV^{-1}day^{-1}}}
\def\mwimp{\rm{m_{\chi}}}

\def\munu{\mu_{\nu}}

\def\sa12{\rm{SA_{12}}}

\begin{document}

\hfill AS-TEXONO/07-05\\
\hspace*{1cm} \hfill \today

\title{Prospects of cold dark matter searches with an
ultra-low-energy germanium detector}

\author{H T Wong$^{1, \ast }$, M. Deniz$^{1,2}$,
H B Li$^1$, S K Lin$^1$, S T Lin$^1$,
J Li$^3$, X Li$^3$, X C Ruan$^4$, 
V Singh$^5$,
Q Yue$^3$, Z Y  Zhou$^4$
}

\address{
$^1$ Institute of Physics, Academia Sinica, Taipei 11529, Taiwan.\\
$^2$ Department of Physics, Middle East Technical University,
Ankara 06531, Turkey.\\
$^3$ Department of Engineering Physics, Tsing Hua University,
Beijing 100084, China.\\
$^4$ Department of Nuclear Physics,
Institute of Atomic Energy, Beijing 102413, China.\\
$^5$ Department of Physics, Banaras Hindu University, 
Varanasi 221005, India. \\
}

\ead{htwong@phys.sinica.edu.tw ($^\ast$Corresponding Author)}

\begin{abstract}

The report describes the research program 
on the development of ultra-low-energy germanium detectors,
with emphasis on WIMP dark matter searches.
A threshold of 100~eV is achieved with 
a 20~g detector array, providing a unique probe 
to the low-mas WIMP.
Present data at a surface laboratory
is expected to give rise to 
comparable sensitivities with 
the existing limits
at the $\rm{ 5 - 10 ~GeV}$ WIMP-mass range.
The projected parameter space to be probed
with a full-scale, kilogram mass-range experiment
is presented.
Such a detector would also allow the studies of 
neutrino-nucleus coherent scattering and
neutrino magnetic moments.
\end{abstract}

Weakly Interacting Massive Particles
(WIMP, denoted by $\chi$) are
the leading candidates for 
cold dark matter (CDM)~\cite{pdgcdm},
and supersymmetric (SUSY) particles
are the favored WIMP candidates.
The popular SUSY models 
prefer WIMP mass ($\mwimp$) of the range
of $\sim$100~GeV,
though light neutralinos remain a possibility~\cite{lightsusy}.
Simple extensions 
of the Standard Model with a singlet scalar
favors light WIMPs~\cite{darkon}.
Most CDM experiments 
optimize their design in the high-mass region, 
and have diminishing sensitivities for $\rm{\mwimp < 10 ~GeV}$,
where there is an allowed region 
if the DAMA annual modulation data 
are interpreted as WIMP signatures~\cite{dama}.
To probe this low-mass region, detectors with sub-keV
threshold are necessary.
These low-threshold detectors will also open the
window to look for WIMPs bound in the solar system~\cite{solarwimp},
as well as non-pointlike SUSY candidates like Q-balls~\cite{qball}.
Sensitivity to sub-keV energy presents a formidable
challenge to detector technology and to background control.
So far, only the CRESST-I experiment have derived
exclusion limits~\cite{cresst1}
with sapphire($\rm{Al_2 O_3}$)-based cryogenic detector
at a threshold of 600~eV.

A research program in low energy neutrino
and astroparticle physics is pursued~\cite{texonoprogram}
by the TEXONO Collaboration
at the Kuo-Sheng(KS) Reactor Laboratory.
A scientific goal is to
develop advanced detectors with
kg-size target mass, 100~eV-range threshold
and low-background specifications
for WIMP searches as well as for
the studies of neutrino-nucleus coherent scatterings~\cite{ulege}
and neutrino magnetic moments ($\munu$)~\cite{munureview}.
The KS laboratory is located 28~m from a 2.9~GW
reactor core and has an overburden of about 
30~meter-water-equivalence~(mwe).
Its facilities are described in Ref.~\cite{texonomagmom},
where the $\munu$-studies with
a 1.06~kg germanium detector (HPGe) at a hardware threshold of
5~keV were reported. This HPGe has also been used
for the studies of reactor electron neutrons~\cite{rnue}
as well as searches for reactor axions~\cite{raxion}.
The experimental procedures were well-established
and the background above 12~keV were measured.
In particular, a background level of about
${\rm \sim 1 ~ event ~ \cpd  }$(cpd)
comparable to those of other
underground CDM experiments
was achieved.

\begin{figure}[hbt]
\begin{minipage}{18pc}
\includegraphics[width=18pc]{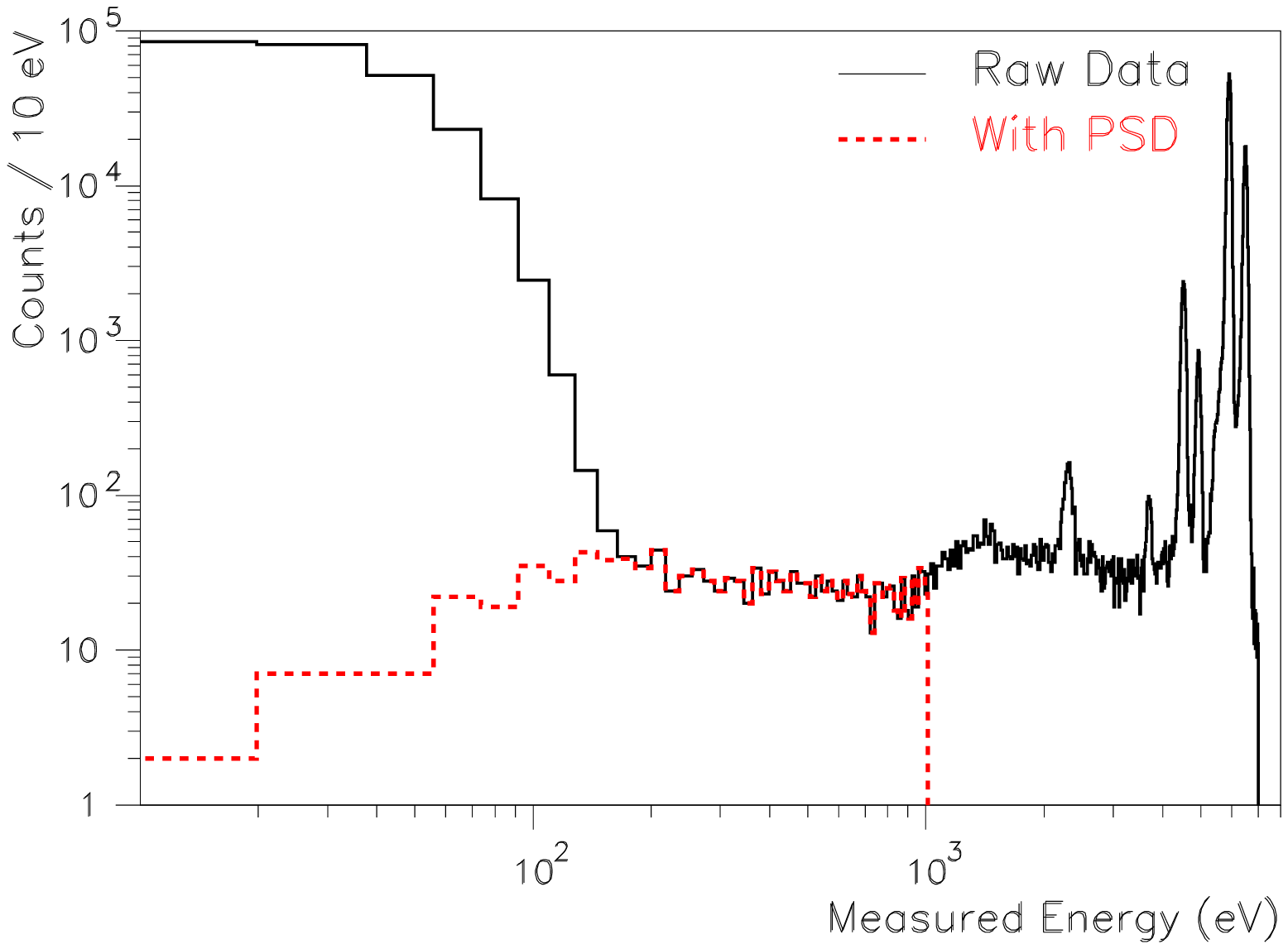}
\caption{\label{fe55}
Measured energy spectra 
with the ULEGe prototype,
using $^{55}$Fe source 
producing X-rays from various isotopes.
A threshold of 100~eV was achieved,
and the electronic noise edge was suppressed
by PSD.
}
\end{minipage}\hspace{2pc}%
\begin{minipage}{18pc}
\includegraphics[width=18pc]{psd.eps}
\caption{\label{psd}
Pulse shape discrimination:
correlations of signals with
different electronic amplifications
and shaping times lead to
suppression of the noise edge.
}
\end{minipage}
\end{figure}

``Ultra-Low-Energy'' germanium (ULEGe)
detectors, developed originally for soft X-rays
detection, are candidate technologies to meet
the challenges of probing into the 
unexplored sub-keV energy domain~\cite{ulege}.
These detectors typically have modular mass of 
5-10~grams. Detector array
of up to N=30 elements have been successfully
built, while there are recent advances in 
developing single-element ULEGe of kg-size mass~\cite{chicago}.
Various prototypes based on this
detector technology have been constructed.
Depicted in Figure~\ref{fe55}
is the measured energy spectrum due to
external $^{55}$Fe calibration sources~(5.90 and  6.49~keV)
together with X-rays from
Ti~(4.51 and 4.93~keV), Ca~(4.01~keV),
S~(2.46~keV) and Al~(1.55~keV). 
Random trigger events uncorrelated to the detector
provided the zero-energy pedestals.
Pulse shape discrimination (PSD)
criteria were applied
as illustrated in Figure~\ref{psd}
by correlating two output with
different electronics amplifications and 
shaping times.
The electronic noise edge
was suppressed by PSD
and a threshold of 100~eV was achieved. 
The deviations of the low
energy spectra of the selected events from
a flat distribution gave the efficiencies
of the PSD cuts.

\begin{figure}[hbt]
\begin{minipage}{18pc}
\includegraphics[width=18pc]{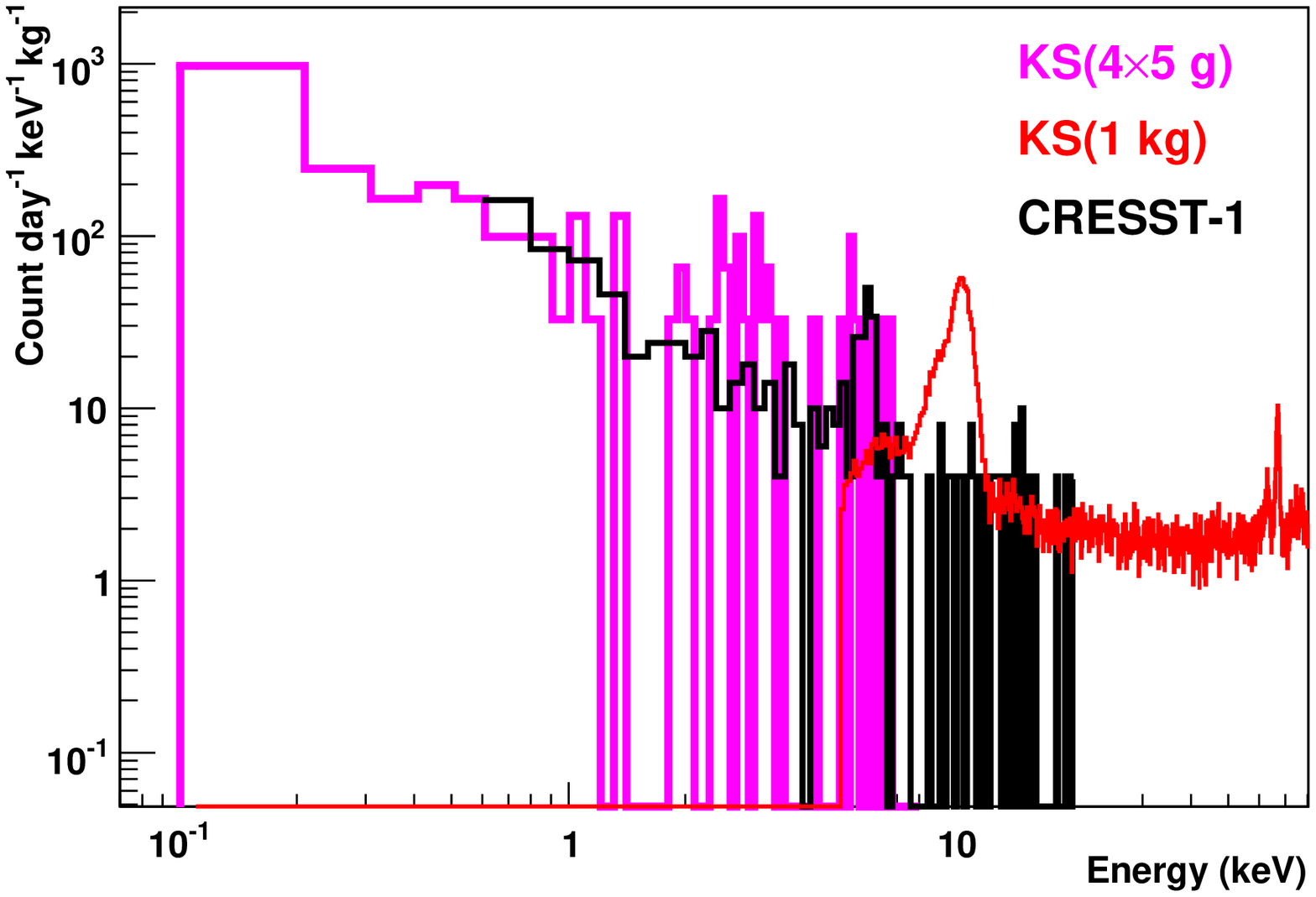}
\caption{\label{bkgspect}
Measured background spectra from 0.34~kg-day of data 
for the KS 4$\times$5~g ULEGe.  Overlaid for comparisons
are those of the KS 1~kg HPGe detector~\cite{texonomagmom} 
and the CRESST-1 experiment~\cite{cresst1}. 
}
\end{minipage}\hspace{2pc}%
\begin{minipage}{18pc}
\includegraphics[width=18pc]{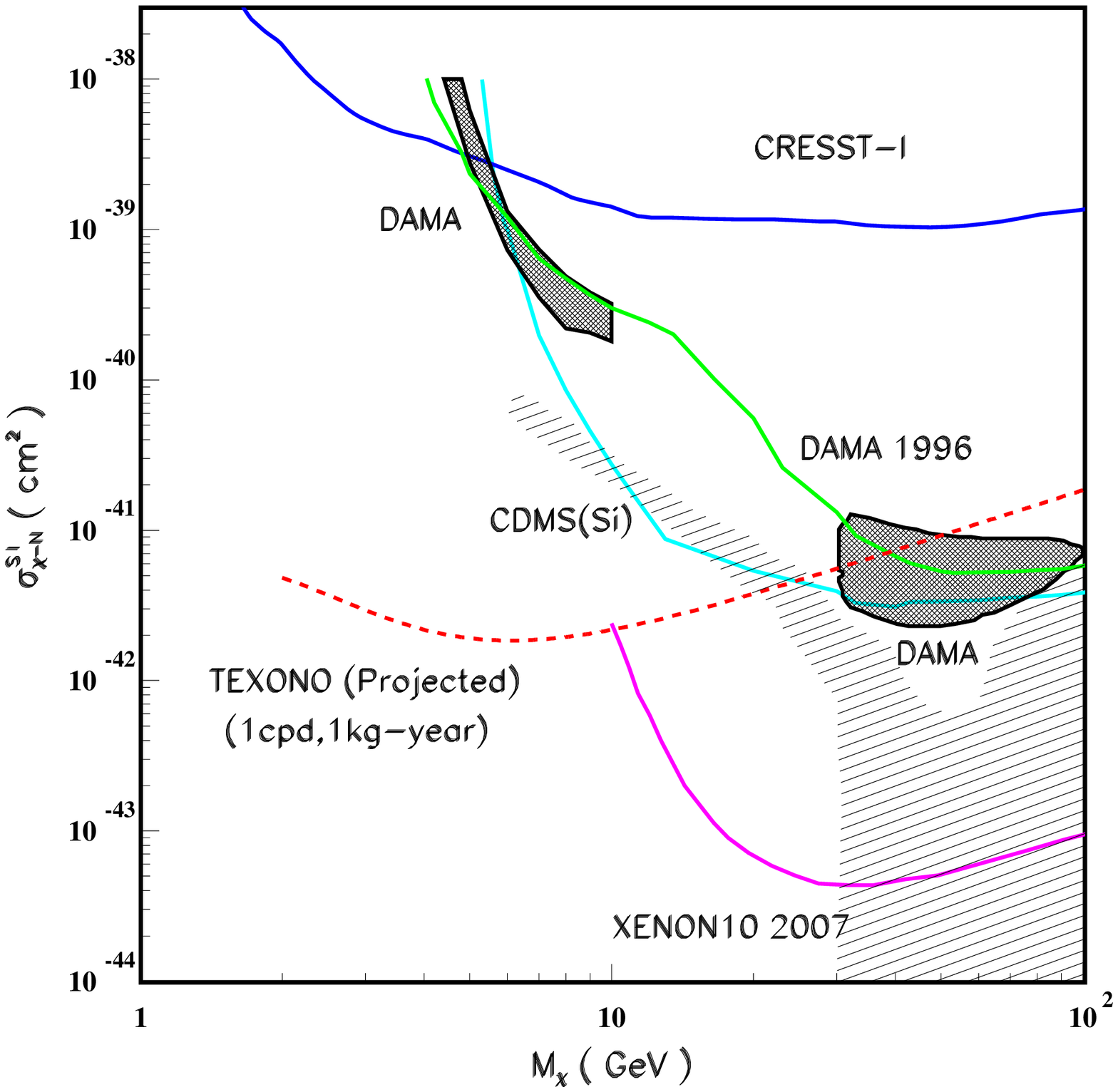}
\caption{\label{dmplot}
Expected spin-independent
sensitivity region for
CDM searches using
a ULEGe detector 
with a mass of 1~kg 
at 1~cpd background level. Also shown are
the present exclusion boundaries~\cite{pdgcdm,taup07}
and the DAMA allowed 
regions~\cite{dama}. The striped region is the  
parameter space favored by the
SUSY models~\cite{lightsusy}.
}
\end{minipage}\hspace{2pc}%
\end{figure}

Low-background data were taken at KS 
under the same shielding configurations as the
magnetic moment experiments~\cite{texonomagmom},
using a 4-channel ULEGe detector each having an
active mass of 5~g.
The recorded spectrum with 0.34~kg-day
of data after cosmic-rays
and anti-Compton vetos and PSD selection
is displayed in Figure~\ref{bkgspect}.
It can be seen that 
comparable background level was achieved as the
CRESST-I~\cite{cresst1} experiment
with 1.51~kg-day of data.
A summary of the spin-independent 
exclusion plot is depicted 
in Figure~\ref{dmplot}.
Comparable limits to CRESST-I
can be expected from the current KS results.
Intensive efforts on the data
analysis are underway.
The projected sensitivities with
1~kg-year of data at 100~eV threshold
and a 1~cpd background level are shown. 

An R\&D program towards the realizations
of a full-scale experiment is rigorously pursued.  
Quenching factor measurement for nuclear recoils in Ge with
sub-keV ionization energy 
will be performed at a neutron beam facility.
Background studies are conducted
at both KS and the Yang-Yang 
Underground Laboratory (700 mwe)
in South Korea with the various prototypes.
Background understanding at the sub-keV range is
a challenging and unexplored subject in its own right.
External background is expected to be reduced
due to self-shielding effects in a kg-mass
detector, as well as by additional 
active veto Ge layers
enclosing hermetically the ULEGe inner target~\cite{ulege}.
Studies are under way with a 180-g 
segmented ULEGe prototype equipped with 
a veto ring and dual-readout channels
from both the signal and high-voltage electrodes.
A 500-g detector similar to the design demonstrated
in Ref.~\cite{chicago} is being constructed.

\end{document}